\documentclass{ieeeaccess}
\usepackage[utf8]{inputenc}
\usepackage{cite}
\usepackage{amsmath,amssymb,amsfonts}
\usepackage{algorithmic}
\usepackage{graphicx}

\usepackage{textcomp}
\def\BibTeX{{\textrm B\kern-.05em{\sc i\kern-.025em b}\kern-.08em
    T\kern-.1667em\lower.7ex\hbox{E}\kern-.125emX}}

\usepackage[dvipsnames]{xcolor}
\usepackage{xspace}

\begin{document}
\doi{10.1109/ACCESS.2026.DOI}

\title{Accurate Data-Based State Estimation from Power Loads Inference in Electric Power Grids}
\author{\uppercase{Philippe Jacquod}\authorrefmark{1,2}, \IEEEmembership{Member, IEEE},
\uppercase{Laurent Pagnier\authorrefmark{3}, \IEEEmembership{Member, IEEE} and Daniel J. Gauthier}\authorrefmark{4},
}
\address[1]{Department of Quantum Matter Physics, University of Geneva, Geneva, Switzerland}
\address[2]{School of Engineering, University of Applied Sciences of Western Switzerland HES-SO, Sion, Switzerland}
\address[3]{Department of Mathematics, The University of Arizona, Tucson, AZ 85721, USA}
\address[4]{ResCon Technologies, 1275 Kinnear Rd., Columbus, OH 43212, USA}
\tfootnote{PJ is grateful to the Andlinger Center for Energy and Environment, Princeton University, for its hospitality during the early stages of this project.}

\markboth
{P. Jacquod, L. Pagnier, D. Gauthier: Accurate Data-Based State Estimation from Power Loads Inference in Electric Power Grids}
{P. Jacquod, L. Pagnier, D. Gauthier: Accurate Data-Based State Estimation from Power Loads Inference in Electric Power Grids}

\corresp{Corresponding author: Philippe Jacquod (e-mail: philippe.jacquod@unige.ch).}

\begin{abstract}
Accurate state estimation is a crucial requirement for the reliable operation and control of electric power systems. Here, we construct a data-driven, numerical method to infer missing power load values in large-scale power grids. Given partial observations of power demands, the method estimates the operational state using a linear regression algorithm, exploiting statistical correlations within synthetic training datasets. We evaluate the performance of the method on three synthetic transmission grid test systems. Numerical experiments demonstrate the high accuracy achieved by the method in reconstructing missing demand values under various operating conditions. We further apply the method to real data for the transmission power grid of Switzerland. Despite the restricted number of observations in this dataset, the method infers missing power loads rather accurately. Furthermore, Newton-Raphson power flow solutions show that deviations between true and inferred values for power loads result in smaller deviations between true and inferred values for flows on power lines. This ensures that the estimated operational state correctly captures potential line contingencies. Overall, our results indicate that simple data-based regression techniques can provide an efficient and reliable alternative for state estimation in modern power grids.
\end{abstract}


\titlepgskip=-15pt

\maketitle

\section{Introduction}
\label{sec:introduction}
Electric power systems are undergoing fundamental changes in their mode of operation. The decarbonization of the energy system requires a transition toward renewable electricity generation, replacing centralized, dispatchable generators based on grid-stabilizing rotating machines, with intermittent, geographically distributed, and inertialess generators~\cite{Smi22,EUC}. Ensuring a safe supply of electricity as well as the stability of present and future power grids with large
penetrations of renewable energy poses significant challenges~\cite{Mil18,Bas20,Den20,Mak21,Ker23}. In particular, power grids increasingly operate closer to their operational limits~\cite{Mart25}, requiring fast and appropriate remedial actions by system operators. To enable rapid response and effective corrective measures, grid operators need to know the state of the system with maximum accuracy. To this end, power grids are being further modernized through smart grid technologies, including advanced sensing and enhanced communication systems for real-time monitoring and control~\cite{Oha24}.

A precise representation of the operating state is the basis on which grid operators analyze the system, evaluate its security and stability, and decide on corrective measures and remedial actions. This typically relies on large, multivariate sets of data, \textit{e.g.}, 
on power injection, voltage amplitude, phase, and frequency at each substation, on the power flow on each line and so forth, acquired and transmitted via the Supervisory Control and Data Acquisition (SCADA) system. 
These datasets are, however, often marred with errors, where data are missing or wrong due to faulty acquisition, erroneous transmission, or following a deception cyber-attack
with false data injection~\cite{Pas13}. Such occurrences will most certainly multiply in the future because the massive deployment of smart grid technologies increases the number of entrance doors to the communication system and the probability that one or a few electronic components malfunction~\cite{Ogi17}. It is imperative to develop algorithmic solutions that detect such data anomalies and estimate the operating state of the system without the knowledge of missing or faulty data. 

Data-based algorithms are natural candidates for anomaly detection and state estimation in the presence of missing or faulty data, regardless of the origin of the anomaly~\cite{Pim14,Thu20}. Owing to their computational simplicity, 
the industry standard for state estimation is weighted least-square methods, where the operating state is guessed as a minimum of a parabolic (convex) cost function~\cite{Dag97,Mon02,Kam08,Thu10}. They effectively find a local minimum
to what is, in principle, a highly nonlinear problem. Not unexpectedly, Artificial Intelligence (AI) algorithms, such as genetic algorithms, particle swarm optimization algorithms, and neural networks, have been proposed to solve such nonlinear problems~\cite{Yan07,Tsai09,Hos09,Pan10,Sun11,Tun16}. 
To the best of our knowledge, earlier works achieve best estimates for state variables, which are voltage magnitude and phases on all buses, i.e. the nodes on the network representing the power grid. Once these variables are known, the state of the system is unambiguously determined.  The flows on each power line is easily computed from the power flow equations, provided the grid topology is known~\cite{bialek}.

In this work, we take a complementary approach by inferring not state variables, but instead active and reactive power demands at load buses, \textit{i.e.}, control variables. Our investigations are based on earlier works, where we constructed a model for the synchronous transmission power grid of continental Europe~\cite{Pag19a,Pag19b,Tyl19}, and generated large sets of statistically realistic data for such power grids, intended to be used for training and testing of 
machine learning (ML) algorithms~\cite{Gil24,Gil25}.
Using these two sets of tools, we train a simple linear regression algorithm, which delivers accurate estimates for power injections at each of the network's thousands of load nodes, even with rather large fractions of missing data. We further apply our algorithm to a size-limited dataset of real historical data for the high voltage transmission grid of Switzerland, where we show that not only the estimation of power injections is accurate, but that, once incorporated into a Newton-Raphson power flow solver, electric power flows on all lines are also estimated accurately. 

Our novel contributions are demonstrating that:
\begin{itemize}
\item state estimation in large electric power grids is successful using a simple, linear regression algorithm;
\item the model infers active power demands at load buses, \textit{i.e.}, control variables, using only available data;
\item the model works for synthetic data representing power grids of different sizes and power generation mixes;
\item the model works for a real power grid with historical data for active power generation. 
\end{itemize}

\section{Notation and Definitions}

The stationary, operational state of a power grid is described by the solution to the power flow equations,
\begin{subequations}\label{eq:powerflow}
\begin{eqnarray}
p_{i} & = & \sum_j v_i v_j [G_{ij} \cos(\theta_i-\theta_j) + B_{ij} \sin(\theta_i-\theta_j)] \, , \qquad \label{eq:active} \\
q_{i} & = & \sum_j v_i v_j [G_{ij} \sin(\theta_i-\theta_j) - B_{ij} \cos(\theta_i-\theta_j) ] \, . \label{eq:reactive}
\end{eqnarray}
\end{subequations}
They connect active and reactive powers $p_i$ and $q_i$ to voltage amplitudes
and phases $v_i$ and $\theta_i$ at the $i=1, \ldots M$ nodes of a network defined by the conductance ($G_{ij}$) and susceptance ($B_{ij}$) matrices modeling the power lines. Active and reactive powers are control variables, while voltage amplitudes and phases are state variables, from which the operational state of the system can be computed. In particular, each term in the summation on the right-hand side of Eqs.~\eqref{eq:powerflow} gives the active [Eq.~\eqref{eq:active}] or
reactive [Eq.~\eqref{eq:reactive}] power flow on the power line connecting 
nodes $i$ and $j$, and whose conductance and susceptance are $-G_{ij}$ and $-B_{ij}$.

Nodes correspond either to generators or loads, depending on the sign of the active power. To differentiate these two types of nodes, we introduce the notation $g_i = p_i >0$ for generators and $l_i = p_i < 0$ for loads. Both types of nodes can inject ($q_i >0$) or consume ($q_i <0$) reactive power. There are $M_\textrm{load}$ loads and $M_\textrm{gen}$ generators, with $M=M_\textrm{load}+M_\textrm{gen}$.

In real life, power grids are never exactly stationary, but evolve slowly as generations are cranked up or down, following increasing or decreasing loads. When considering the operational state at different times, a timestamp needs to be introduced, and below we will consider sequences of active power loads, $l_{i,\tau}$, and generations, $g_{i,\tau}$, at discrete time intervals labeled by $\tau=1, \ldots N$, where $N$ gives the number of observation times.  
 


Earlier methods focusing on the state variables $\{v_i,\theta_i\}$ directly give informations on line currents, power flows and possible line contingencies, voltage drops and so forth, which are of central importance for grid operators. The state evaluation to be presented below is based instead on the  control variables $\{p_i,q_i\}$. It is therefore important to translate deviations of predicted vs. true power variables into deviations of predicted vs. true state variables. Here we do this under the standard assumption that susceptance terms dominate over conductance terms at high voltage levels, $|B_{ij}| \gg |G_{ij}|$.

Linearizing the right-hand side of Eq.~\eqref{eq:active} in voltage angle deviations $\theta_i - \theta_i^{(0)}= \Delta \theta_i$
between true $\theta_i^{(0)}$ and inferred $\theta_i$ voltage angles, 
we obtain
\begin{subequations}
\begin{eqnarray}
\Delta p_i & = & \sum_j \Delta p_{ij} \, , \\
\label{eq:dpij}
\Delta p_{ij} &=& B_{ij} v_i v_j \cos(\theta_i^{(0)}-\theta_j^{(0)}) (\Delta \theta_i - \Delta \theta_j) \, , 
\end{eqnarray}
\end{subequations}
which gives a linear relation between power and voltage angle deviations. 
Note that 
similar linear relations
between $\{p_i,q_i\}$ and $\{v_i,\theta_i\}$ can be obtained by linearizing
Eqs.\eqref{eq:active} and \eqref{eq:reactive} in both $\Delta \theta_i$ and $\Delta v_i$,
even relaxing the simplifying assumption $|B_{ij}| \gg |G_{ij}|$.

It is obvious from Eq.~\eqref{eq:dpij} that
power flow deviations $\Delta p_{ij}$ can be positive or negative, depending on the sign of 
$\Delta \theta_i - \Delta \theta_j$. We therefore write $\Delta p_{ij}=\overline{\Delta p_i} + \delta p_{ij}$ as the sum of a finite (positive or negative) average, $\overline{\Delta p_i} = c_i^{-1} \sum_j \Delta p_{ij}$ and a fluctuating term with
zero average, $\sum_j \delta p_{ij}=0$, with the connectivity $c_i$ giving the number of power lines
connected to the $i^{\rm th}$ node.

Using the linearized model, we find 
$\Delta p_{i}^2 = c_i^2 \, \overline{\Delta p_i}^2 + c_i \, \rm{ Var} (\delta p_{ij})$. 
If power flow deviations are dominated by their average, then they are smaller than power injection deviations by a factor proportional to the node connectivity $c_i$ so that $\Delta p_{ij} \simeq \Delta p_i/c_i$, whereas the reduction factor is proportional to the 
square root of the connectivity $|\Delta p_{ij}| \simeq |\Delta p_i|/\sqrt{c_i}$ if their fluctuations dominate. Either way, we conclude that small errors in inferring power injections should result in smaller errors in power flows. Furthermore, the error $\Delta p_{ij}$ may {\it a priori} go either way, increasing or decreasing the power flow, \textit{i.e.}, $|p_{ij} +\Delta p_{ij}| = |p_{ij}| \pm |\Delta p_{ij}|$, so that deviations may predict either more or less loaded power lines. 

\section{Inference Model}\label{section:model}

In this section, we briefly outline the model for inferring power grid loads and methods for finding the model parameters.  For each power grid observation $\tau=1,\ldots N$, we have a set of loads and generator values
\begin{eqnarray}
    \ell_{i,\tau}\hspace{0.1in}&&i=1,\ldots, M_\textrm{load} \, , \nonumber \\
    g_{i,\tau}\hspace{0.1in}&&i=1,\ldots, M_\textrm{gen} \, . \label{eq:load_def}
\end{eqnarray}

In some studies, we remove the $M^\ell_\textrm{top}$ top loads -- those corresponding to the largest active power drawn from the grid -- from the dataset and use the remaining $M^\ell=M_\textrm{load}-M^\ell_\textrm{top}$ loads to infer the loads left out.  The removed load data is gathered in a row vector
\begin{equation}
y^\ell_\tau = [\ell_{M^\ell+1,\tau},\ell_{M^\ell+2,\tau},\ldots,\ell_{M_\textrm{loads},\tau}] \label{eq:y_j}
\end{equation}
of dimension $(1 \times M^\ell_\textrm{top})$. The remaining load data is gathered in a row vector
\begin{equation}
x^\ell_\tau = [\ell_{1,\tau},\ell_{2,\tau},\ldots,\ell_{M^\ell,\tau},1], \label{eq:x_j}
\end{equation}
of dimension $(1\times (M^\ell+1))$. The row vector is augmented by a component equal to 1 at the end, which will be discussed below.  In the machine learning community, $x^\ell_\tau$ is called the `feature vector.' 

The observations are vertically concatenated to form observation matrices $\mathbf{X}^\ell$ (dimension [$N\times (M^\ell$+1)]) and $\mathbf{Y}^\ell$ (dimension $(N \times M_\textrm{top}^\ell)$), where bold indicates a matrix.  
Our goal is to infer the left-out loads at each observation using the remaining load data.  We adopt a linear model given by
\begin{equation}
\mathbf{Y}^\ell = \mathbf{X}^\ell\mathbf{W}^\ell. \label{eq:full_model}
\end{equation}
where the weight matrix $\mathbf{W}$ 
has dimension $((M^\ell+1)\times M_\textrm{top}^\ell)$.  The bias (or offset) of the linear model is taken into account by the 1 appearing in the last element of the row vector in Eq.~\eqref{eq:x_j} and the bottom row of $\mathbf{W}^\ell$. While it is possible to use more complex models, our philosophy is to start with the simplest model and only increase its complexity as needed. Our task is to identify 
a single matrix $\mathbf{W}^\ell$ that works all observations.

We use a data-driven approach for model identification known as supervised learning. Here, we use a subset of the observations to `train' the model and then use the remaining observations to `test' the model, where we use the machine learning community nomenclature.  The test samples have never been `seen' by the model during training, and the identified model is used to predict loads of future observations, also never seen by the model.  This approach assumes that the training dataset is representative of the underlying probability distribution of the data.  

Equation $\eqref{eq:full_model}$ can be solved using regularized least-squares regression, often referred to as ridge regression or Tikhonov regularization.  This method obtains the weights through the relation
\begin{equation}
\mathbf{W}^\ell = [(\mathbf{X}^\ell)^T\mathbf{X}^\ell+\alpha\mathbf{I}]^{-1}(\mathbf{X}^\ell)^T \mathbf{Y}^\ell, \label{eq:W}
\end{equation}
where $\alpha$ is the ridge regression parameter, $T$ is the matrix transpose, and $\mathbf{I}$ is the $((M^\ell+1)\times(M^\ell+1))$ identity matrix.

Regularization achieves several equivalent operations.  It puts a floor $\alpha$ on the singular values of the data covariance matrix $(\mathbf{X}^\ell)^T\mathbf{X}^\ell$ to improve the numerical stability of the algorithm used to find the matrix inverse in Eq.~\eqref{eq:W}.  Equivalently, it penalizes the $L_2$ norm of $\mathbf{W}^\ell$, ensuring that the individual matrix elements do not take on large values.  Equivalently, it prevents the model from over-fitting to the training data and thus improves its generalization to data it has not seen.

In our studies, we find the regularized solution to Eq.~\eqref{eq:full_model} using the Python machine learning routine \texttt{Ridge} that is part of the \texttt{sci-kit learn} package \cite{ridge}, version 1.7.2.  This routine uses singular-value decomposition for the matrix inverse, which is known to give accurate results for similar tasks \cite{Bollt2025}.

We also consider predicting the generator values.  Here, we remove the $M^g_\textrm{top}$ top generators.  The $j^{th}$ observations of all the loads and the remaining $M^g = M_\textrm{gen}-M^g_\textrm{top}$ generators are gathered in a row vector $x^g_j$ (dimension $(1 \times (M^g+M_\textrm{load}+1))$) analogous to Eq.~\eqref{eq:x_j}. The removed top generators are gathered in a row vector $y^g_j$.  The observations and removed loads are vertically stacked in matrices $\mathbf{X}^g$ and $\mathbf{Y}^g$, respectively, and the linear model is given by $\mathbf{Y}^g = \mathbf{X}^g\mathbf{W}^g$.

\section{Datasets}
The synthetic datasets are obtained using the method described in Refs.~\cite{Gil24,Gil25}. They correspond to three different transmission grids coupling
the 220 and 380 kV levels. They give active power for 

$\bullet$ 163 loads and 36 generators for the Swiss power grid,

$\bullet$ 908 loads and 61 generators for the Spanish power grid,

$\bullet$ 560 loads and 101 generators for the German power grid.

For each grid, these synthetic datasets correspond to 20 years with hourly resolution, for a total of 174,720 samples (each synthetic year corresponding to exactly 52 weeks, i.e. 364 days). Values for each generator or load at each time step correspond to 
active power in per unit, corresponding to a base unit of 100 MW. 
Below, we use 80\% of those datasets as training sets and the remaining 20\% for testing the learned model.

The experimental data were provided by the Swiss transmission system operator, Swissgrid, and consist of readings from every Swiss substation for the months of January and July 2015, with a 15-minute time resolution. Substations equipped with transformers usually have two readings, one at each voltage level; additionally, measurements may be split across different busbars. This splitting is exogenous, as it depends on the operator’s actions, and is therefore highly predictable by the operator. We consequently consolidate busbar-level quantities into voltage-level quantities. We also remove trivial recordings, as they provide no useful information and would otherwise hinder the analysis. The system consists of 161 buses, 31 international and 244 national lines.

\section{Results}
\subsection{Inferring The Largest Five Loads}\label{sec:synthetic}
In this section, we study the prediction accuracy when either of the five loads with largest active
power demand are removed from the training dataset, and the remaining loads are used to predict them during model testing. 
We focus on these five largest loads, because failure to evaluate them accurately would in general generate the largest errors in state estimation.
For each dataset, we randomly select 80\% of the data for training the model using Eq.~\eqref{eq:W}, and use the trained model to predict the remaining data during the testing phase.

\begin{figure}[ht!]
\includegraphics[width=0.9\columnwidth]{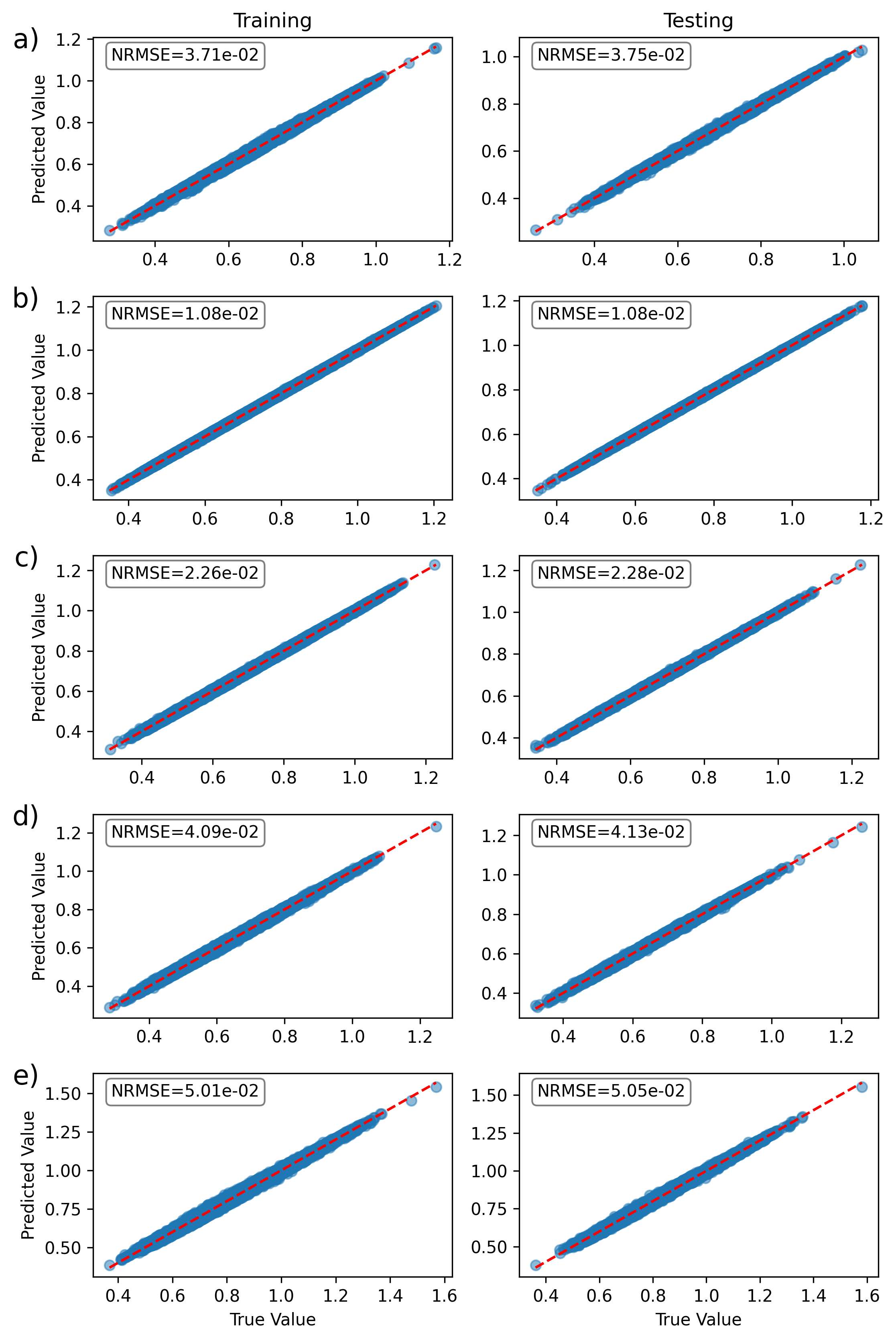}
\caption{Model performance for the synthetic Swiss power grid. Panels a)-e) are predictions for the largest five loads in descending order.}
\label{fig:CH_composite}
\end{figure}

\begin{figure}[ht!]
\centerline{\includegraphics[width=0.9\columnwidth]{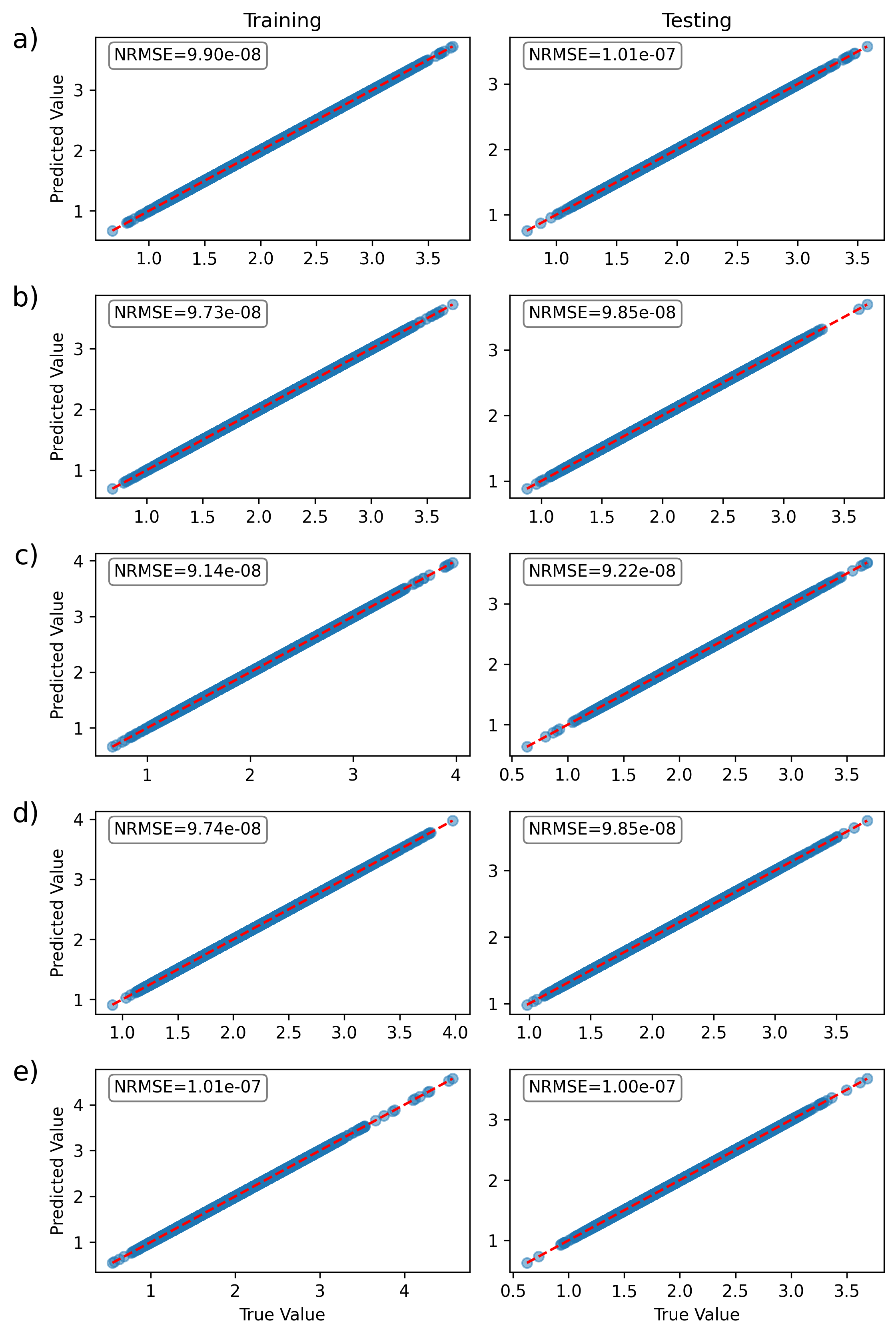}}
\caption{Model performance for the synthetic Spanish power grid. Panels a)-e) are predictions for the largest five loads in descending order.}
\label{fig:ES_composite}
\end{figure}

\begin{figure}[ht!]
\centerline{\includegraphics[width=0.9\columnwidth]{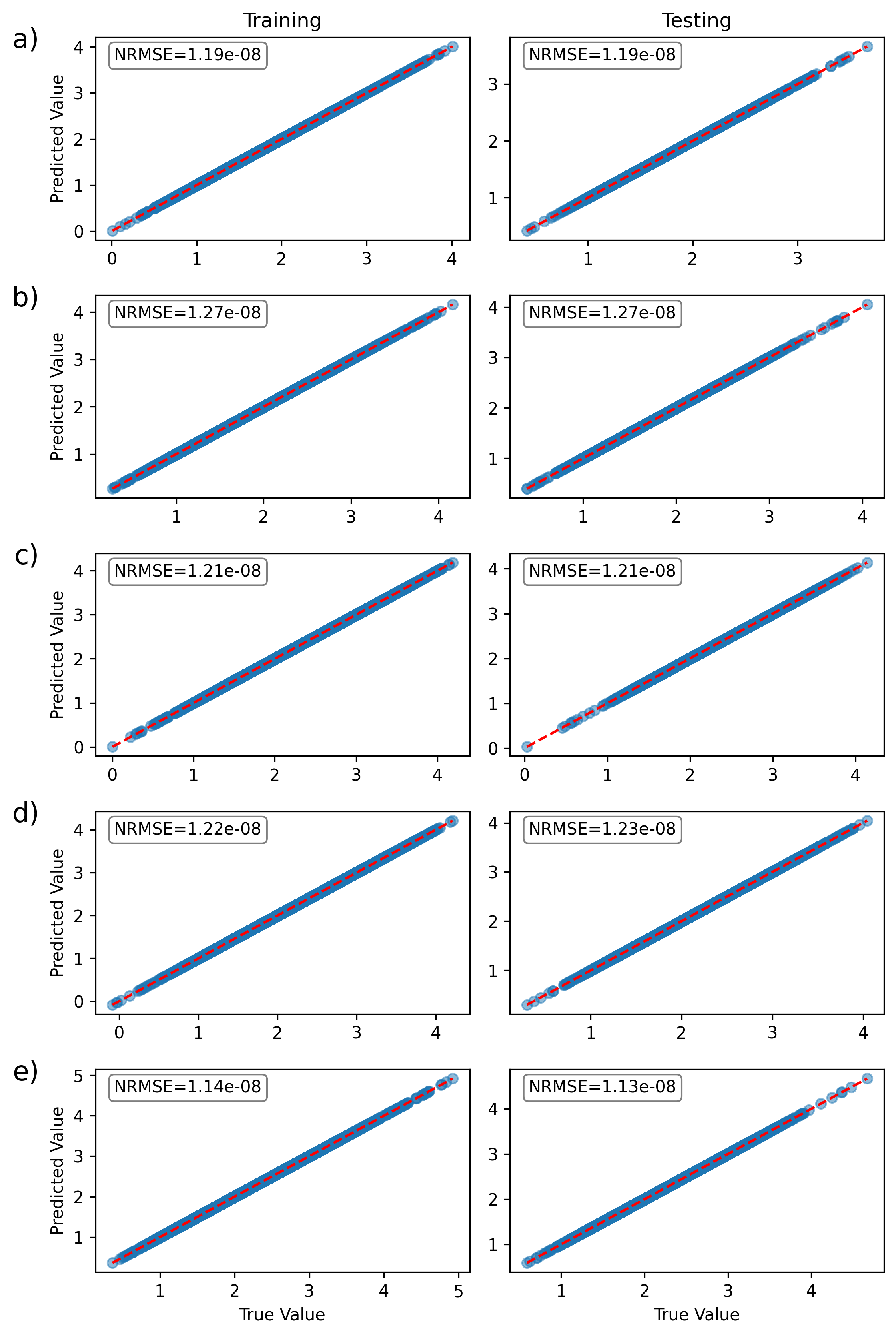}}
\caption{Model performance for the synthetic German power grid. Panels a)-e) are predictions for the largest five loads in descending order. }
\label{fig:DE_composite}
\end{figure}

Figures \ref{fig:CH_composite}, \ref{fig:ES_composite}, and \ref{fig:DE_composite} show the training and testing predicted loads as a function of the true values for the synthetic Swiss, Spanish, and Germany grids, respectively, for $\alpha = 1\times10^{-5}$.  The normalized root mean square error (NRMSE) is given in each panel.  Here, the NRMSE is the root-mean-square error normalized by the standard deviation of each load distribution taken from the training dataset.

For all cases, the true and predicted results are highly correlated for the training and testing data, with the points clustering along the unit-slope line.  The NRMSE is below $\sim 5\times 10^{-2}$ for the Swiss grid, and below $\sim 1\times 10^{-7}$ for the Spanish and German grids.  Importantly, the NRMSE for the test samples is similar to that for the training samples, indicating good model generalization to data not seen during training.

For the discussion in Sec.~\ref{sec:discussion}, we give the model training time for the task described in this section using an Intel 11$^{th}$ Gen Core i7-1165G7 central processor unit with 16 GB random access memory.  For the largest dataset (Spanish grid), there are 908 loads, and we remove the top 5 loads, resulting in 903 features and 139,776 observations (80\% of the total dataset). The regularized regression takes 8.26 s. Applying the trained model to predict the training dataset, used to find the training error, takes 364 ms, or 2.08 $\mu$s per prediction (inference).

\subsection{Model performance for different $M$}

\begin{figure}[ht!]
\centerline{\includegraphics[width=0.9\columnwidth]{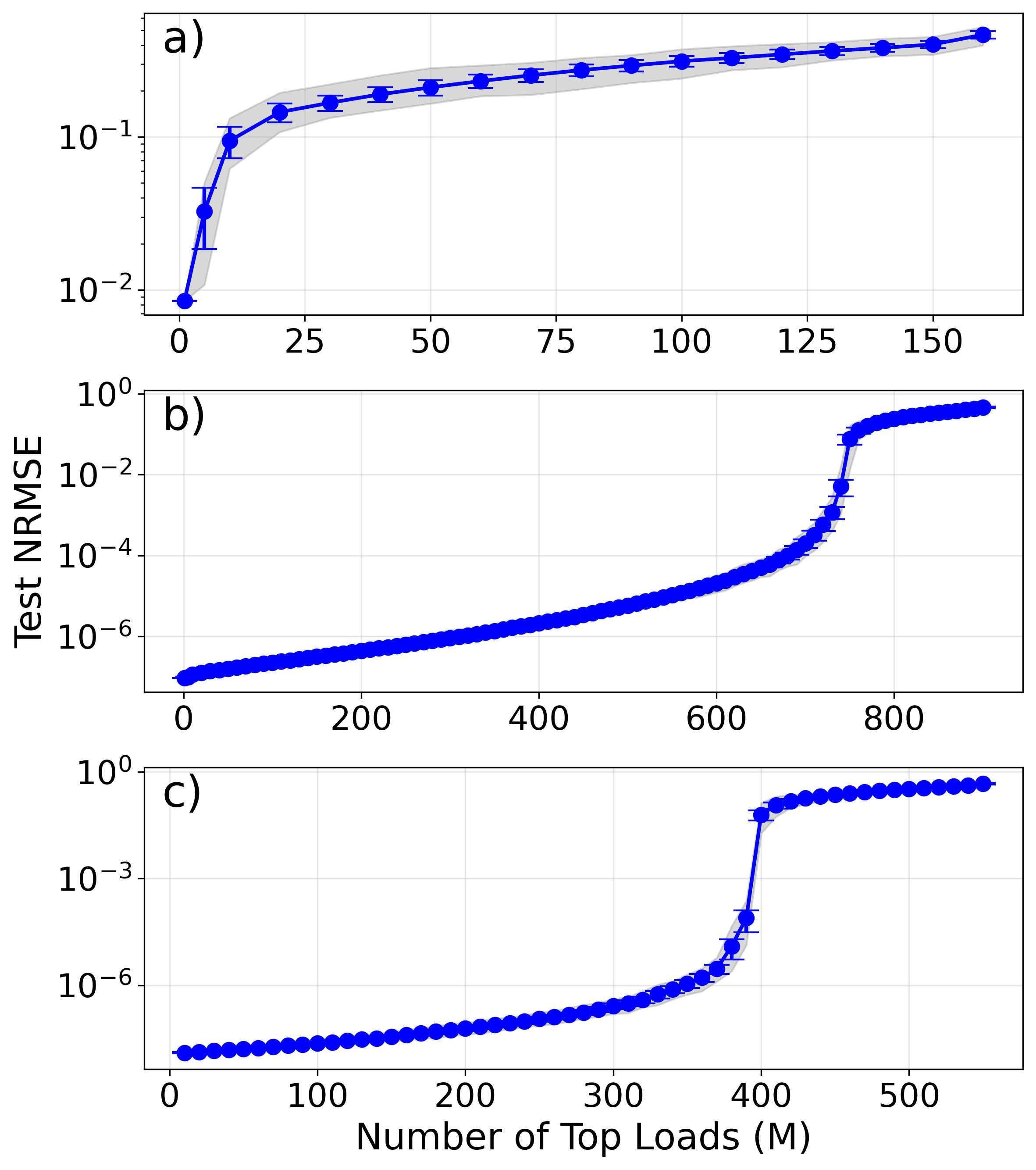}}
\caption{Model performance for inferring the loads with largest active power demand for the synthetic a) Swiss, b) Spanish, and c) German grids, as a function of the number $M$ of left-out loads.}
\label{fig:performance_n_loads}
\end{figure}

To push the model, we investigate the model performance, quantified by the test NRMSE as a function of the number of left-out loads $M$.  As seen in Fig.~\ref{fig:performance_n_loads}a), the Swiss grid performance degrades rapidly with increasing $M$, where the performance rolls over at $M\sim12$.  On the other hand, the Spanish (Fig.~\ref{fig:performance_n_loads}b)) and German (Fig.~\ref{fig:performance_n_loads}c)) grids have very good performance, but degrades approximately exponentially with $M$ for $M<700$ ($M<400$) for the Spanish (German) grids.  Beyond this value, there is a rapid transition to high error for larger $M$.

\subsection{Model Performance for different training set sizes}

\begin{figure}[ht!]
\centerline{\includegraphics[width=0.9\columnwidth]{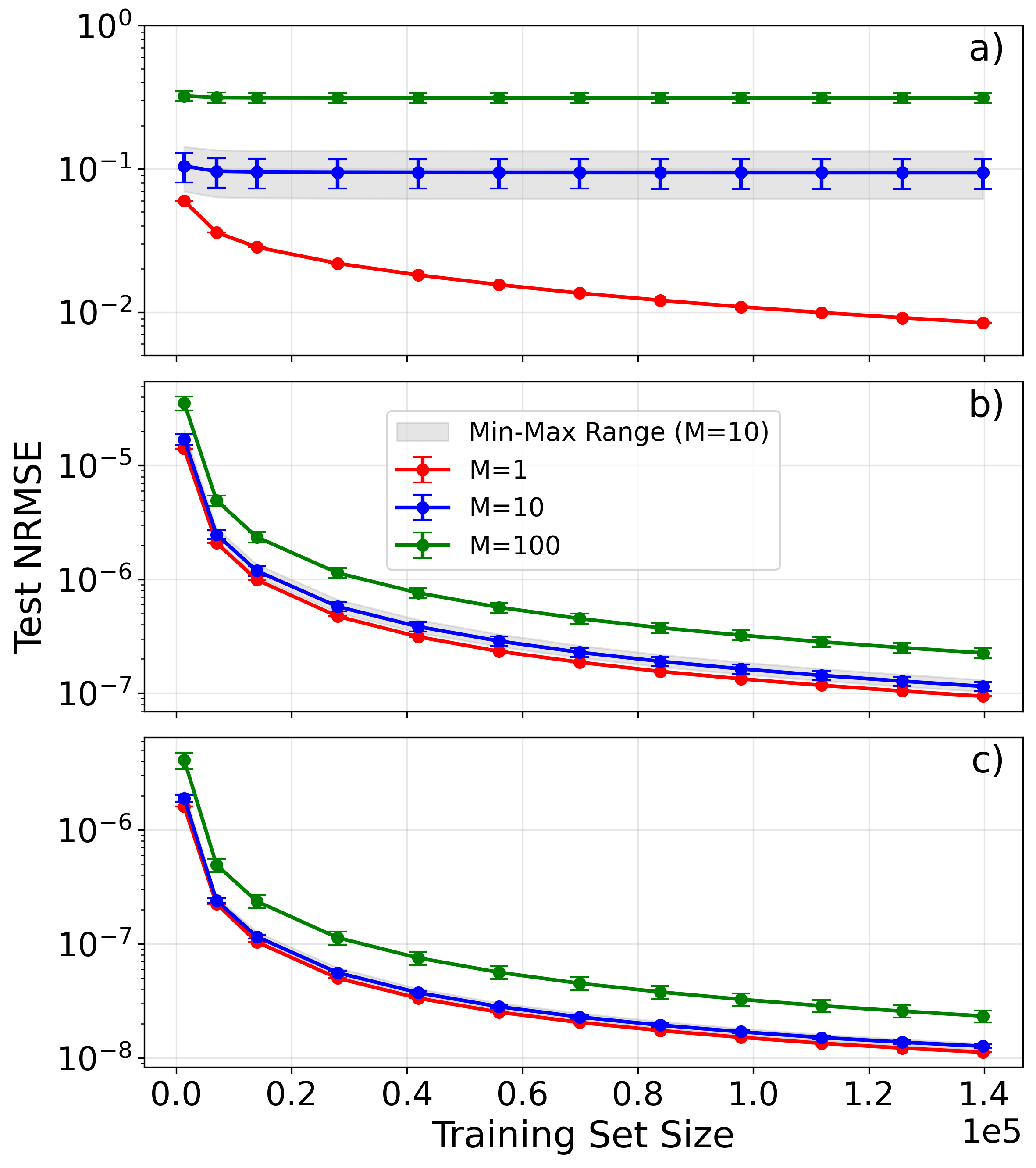}}
\caption{Model performance for different training dataset sizes for the synthetic a) Swiss, b) Spanish, and c) German  grids for different values of $M$.  The gray band for $M=10$ corresponds to the minimum and maximum prediction accuracy.}
\label{fig:performance_train_size}
\end{figure}

An important practical question is the amount of training data required to obtain a given model error.  Figure~\ref{fig:performance_train_size} shows the NRMSE as a function of the training set size for the three synthetic grids.  Here, we first perform the random selection of the training and testing data points, and then we use a fraction of the training data for model determination.  Specifically, the same testing data is used in all cases.  The smallest training dataset size considered is 1,398 samples (1\% of the training set size).

For $M=1$ (red lines) and all country grids, the error decreases continuously as the model is provided with more data.  A prediction error of 1\% can be obtained for the Swiss grid for $\sim 1.1\times10^5$ training data points, whereas the minimum set size considered here (1,389) is enough for the Spanish and German grids.

For $M=5$ or 10, the performance for the Swiss grid is poor and does not improve with greater training data.  On the other hand, the error is nearly identical for $M=1$ and 5 for the Spanish and German grids, and is somewhat higher for $M=100$.  In all cases, the curves follow a similar pattern.

\subsection{Model Weights}

\begin{figure}[ht!]
\centerline{\includegraphics[width=0.7\columnwidth]{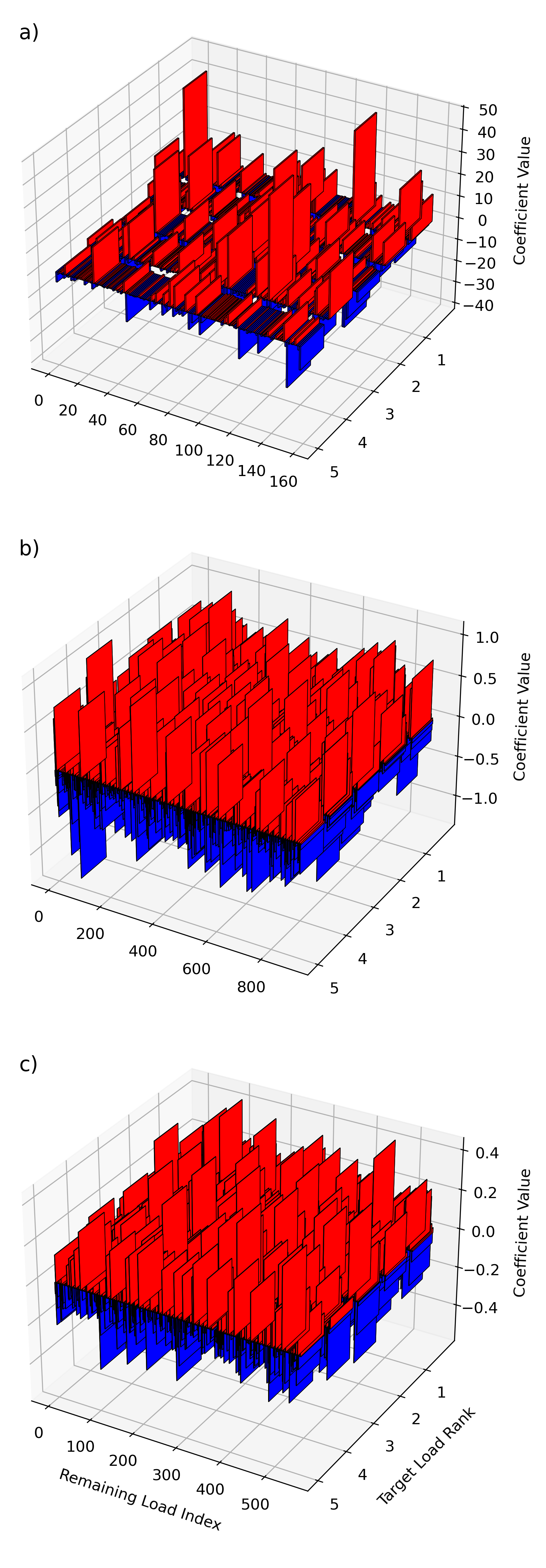}}
\caption{Model interpretation. Model weights for the synthetic a) Swiss, b) Spanish, and c) German grid.  Positive (negative) weights are shown as red (blue), and weights < 0.1 are not shown for clarity.}
\label{fig:combined_3d_model_weights}
\end{figure}

One important advantage of our linear model is that the model weights are proportional to the importance of a load in predicting the left-out loads. Figure~\ref{fig:combined_3d_model_weights} shows the weights for the case when the top 5 loads are left out.  It is seen that the Swiss grid has a small number of important nodes (large weight magnitude), whereas the weights are all important for the Spanish and German grids. This may explain the relatively poor performance of the Swiss grid for more than a few left-out nodes as seen in Fig.~\ref{fig:performance_n_loads}. 

\begin{figure}[ht!]
\centerline{\includegraphics[width=0.7\columnwidth]{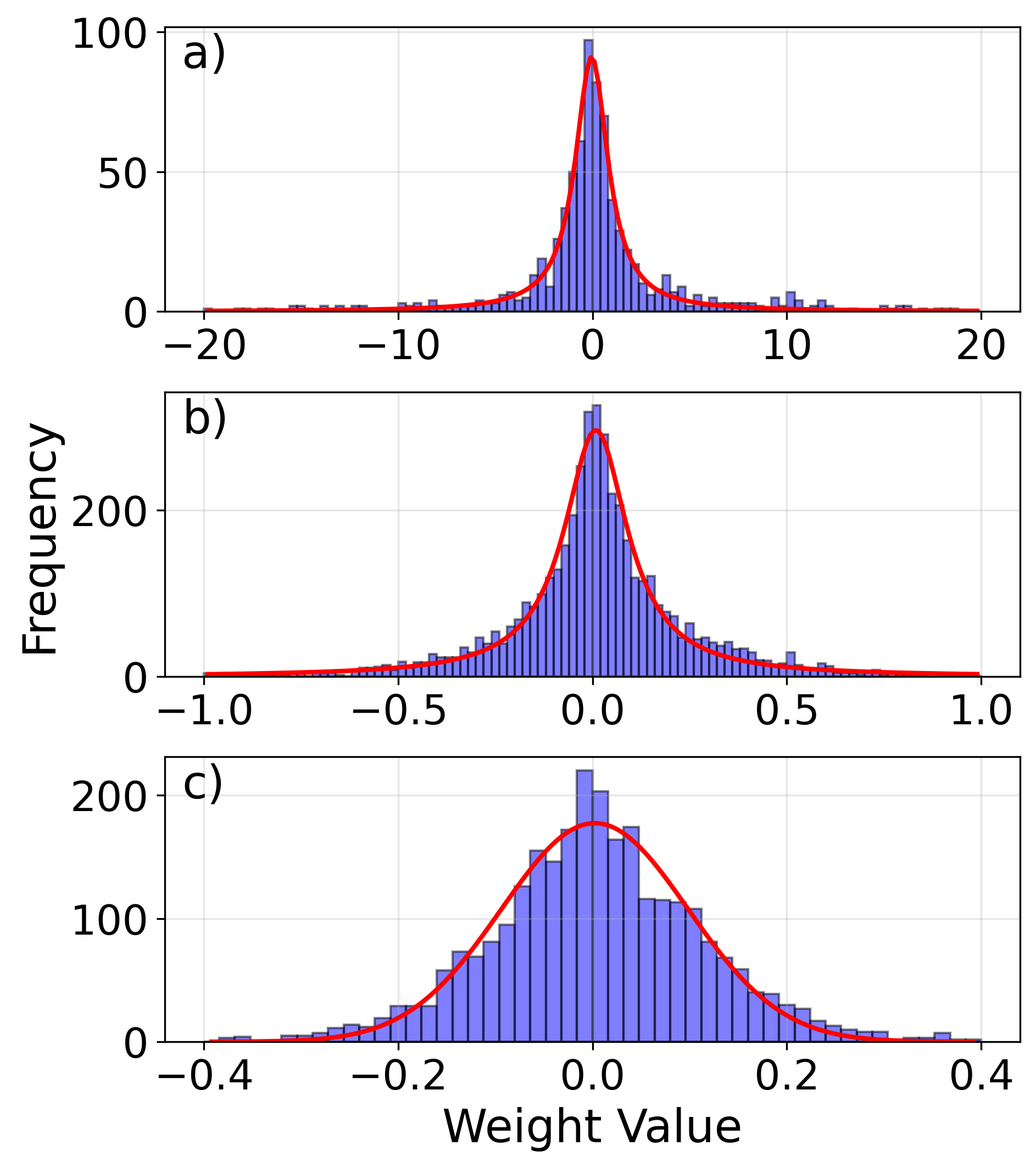}}
\caption{Quantitative model interpretation. Histogram of the model weights for the synthetic a) Swiss, b) Spanish, and c) German grid.  The data is fit to either a Gaussian function $a \exp(-(x-x_0)^2/\sigma^2)$ or a Lorentzian function $a\gamma^2/(\gamma^2+(x-x_0)^2)$.  The best-fit functions are: a) Lorentzian with $a$ = 91.1, $x_0 = -0.042$, $\gamma = 1.03$, and a reduced $\chi_r^2$ = 0.77; b)  Lorentzian with $a$ = 296., $x_0$ = 0.0087, $\gamma$ = 3.55, and $\chi_r^2$ = 3.5; and c) Gaussian with $a$= 177., $x_0$ = 0.00237, $\sigma$ = 0.0961, and $\chi_r^2$ = 2.52.}
\label{fig:combined_weights_histogram}
\end{figure}

The 3d plots indicate that there are many small weights for the synthetic Swiss grid compared to the other two.  To help make a quantitative statement about the weights, we plot the weight histogram shown in Fig.~\ref{fig:combined_weights_histogram}.  We fit the distribution to Gaussian and Lorentzian functions and keep the distribution with a reduced chi-square function ($\chi_r^2)$ closest to one.  Here, the variance is taken as the square root of the number of observations.

We find that the model weight distributions for the Swiss and Spanish grids are best fit by the Lorentzian, whereas the German grid is best fit by a Gaussian.  Notably, the Swiss grid has several small weights that extend beyond the plot boundary, supporting our previous conclusion that there are a few large weights that dominate the model for the Swiss grid in comparison to the other two.

\subsection{Predicting Real Loads and Inferring Power Flows}

The above results have been obtained using synthetic
datasets~\cite{Gil24,Gil25} generated by an
algorithm that preserves the degree of correlation observed
in real, but size-limited historical datasets for the swiss
high voltage power grid. The latter provide power injections and loads for the months of January and July 2015, with an hourly resolution. Although likely too small for training our algorithms, we nevertheless apply our method to these size-limited but real-world datasets. 

\begin{figure}[ht!]
\centerline{\includegraphics[width=0.9\columnwidth]{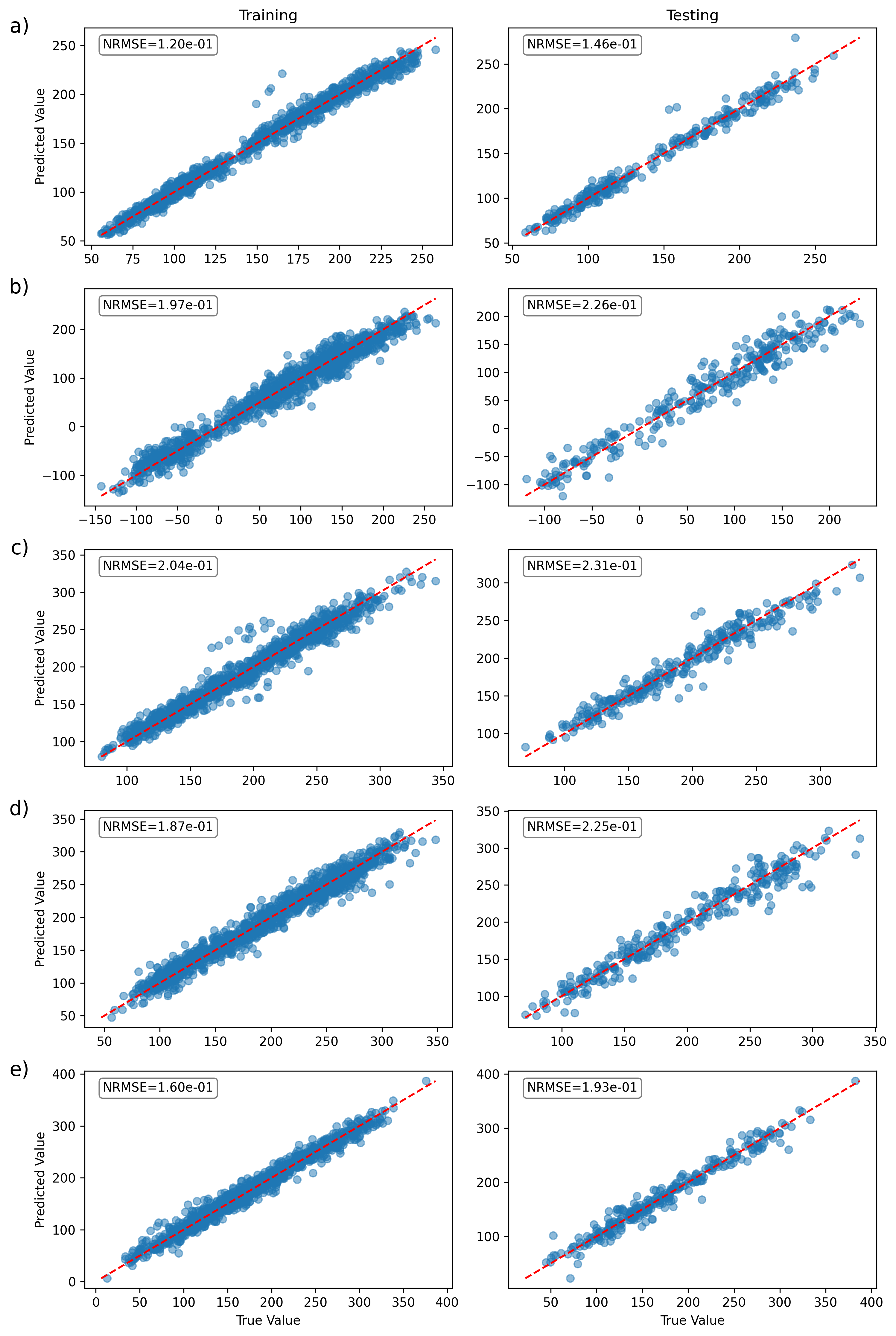}}
\caption{Model performance on real data for the swiss
high voltage power grid. Panels a)-e) are compare predicted with true loads, for the five largest  loads in descending order.}
\label{fig:Swiss_composite}
\end{figure}

The result is shown in Fig.~\ref{fig:Swiss_composite}. We observe a high correlation between true and predicted loads, for the five largest loads in the power system, despite the smallness of the training dataset.  The NRMSE is $\sim 0.25$ for all loads, which is significantly higher than the above  performances for the synthetic data.  This is expected because we have 10$\times$ fewer training samples compared to the synthetic datasets, and there is noise in any real dataset.  To improve the prediction performance for this model, we set $\alpha = 10^{4}$ found using a grid search. 

\begin{figure}[ht!]
\centerline{\includegraphics[width=0.7\columnwidth]{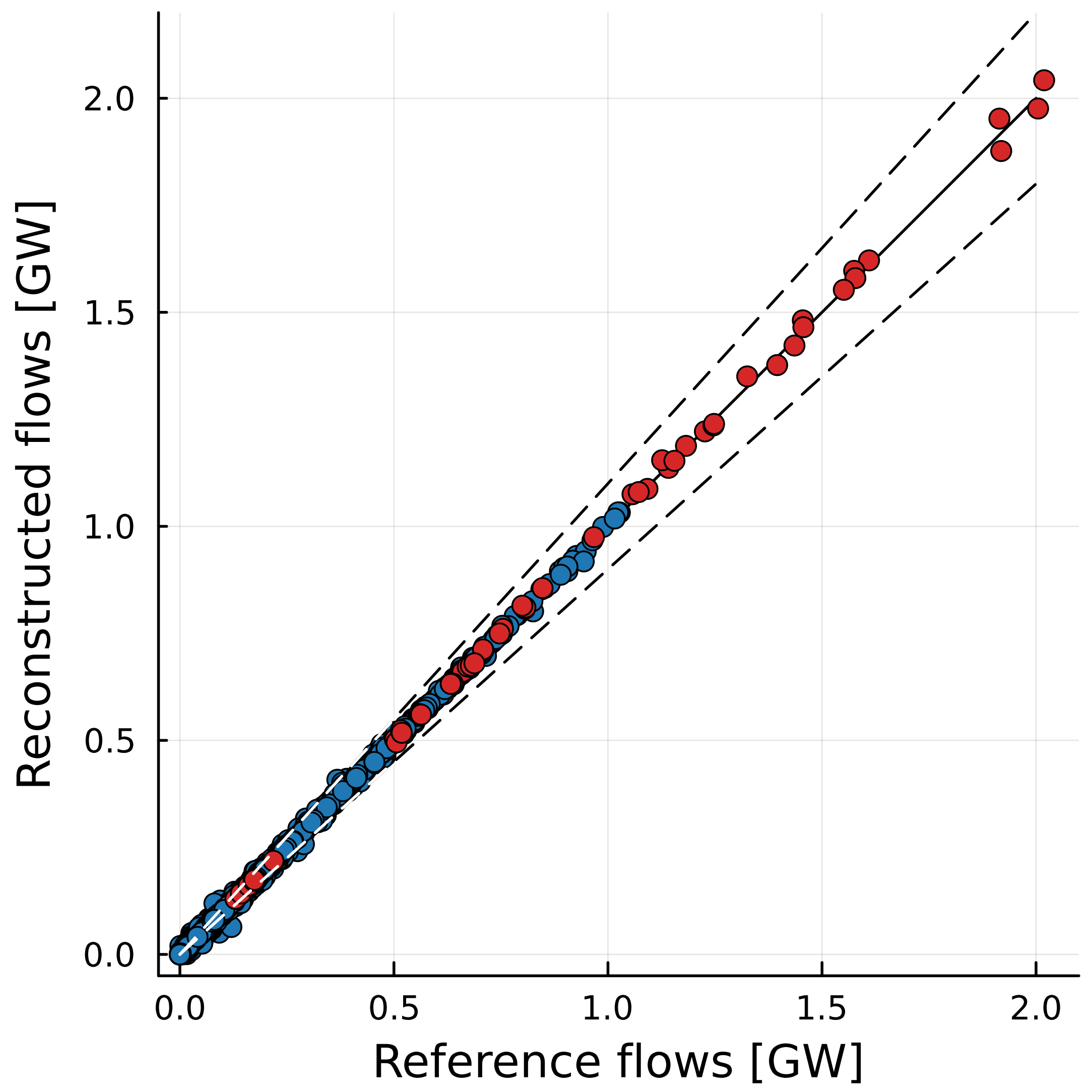}}
\caption{Model performance on real data for the Swiss high voltage power grid: comparison of true and reconstructed flows obtained by Newton- Raphson solutions of the power flow Eqs.~\eqref{eq:powerflow} with
data from Fig.~\ref{fig:Swiss_composite}. Blue (red) dots correspond to power flows below (above) 95 \% of the thermal power. dashed lines indicate deviations of $\pm 10$\%.}
\label{fig:Swiss_flows}
\end{figure}

So far, we focused on inferring power loads. From the point of view of a grid operator, however, other quantities such as voltage drops, current flows and  power flows on lines are of higher importance, as they determine the presence or absence of contingencies that may jeopardize grid stability and the safety of power supply. 
We want to determine whether the inaccuracies inherent to our method can be tolerated in the sense
that: (i) the predicted flows are not too different from the real ones; and (ii) significant deviations between predicted and real flows do not systematically affect already strongly loaded lines. 
To that end, we numerically infer flows on power lines from power injections by solving Eq.~\eqref{eq:powerflow} using 
a Newton-Raphson solver for both predicted and true power injection data for the Swiss transmission power grid. The result is shown in Fig.~\ref{fig:Swiss_flows}. It is seen that deviations rarely exceed 10 \%, and that this does not systematically affect more heavily loaded lines. 

To quantify the accuracy of power flow reconstruction, we introduce three different metrics, which we briefly describe.
From Eq.~\eqref{eq:active},
the real active power flow on the power line connecting buses $i$ and $j$ is
\begin{equation}
p_{ij} = v_iv_j \big[G_{ij}\cos(\theta_i -\theta_j)+ B_{ij}\sin(\theta_i -\theta_j)\big]\,.
\end{equation}
The statistical error in our estimation of the flow on that line can be quantified by the variance
\begin{equation}\label{eq:var}
\rm{var}[\, p_{ij} \,] = N^{-1} \sum_{\tau=1}^N [p_{ij}(\tau) -\hat{p}_{ij}(\tau)]^2 \, , 
\end{equation}
over the testing set with $N$ observation times,
and $\hat{p}_{ij}$ gives the active power flows obtained from 
Eq.~\eqref{eq:active} with the predicted values of $p_i$ from our evaluation algorithm. The accuracy of the predicted power flow solution can then be measured by averaging over all power lines. This can be done in two ways, either by extending  Eq.~\eqref{eq:var} to a global variance calculated over
all power lines and taking the square root of the result, or by first taking the square root of Eq.~\eqref{eq:var} and averaging over all power lines. In both instances, one normalizes the result with the variance $\sigma_{ij} \equiv  N^{-1} \sum_{\tau=1}^N \big(p_{ij}(\tau) -\langle p_{ij}(\tau)\rangle \big)^2$
of the true power flows, calculated over all observation times. One obtains the following two metrics,
\begin{align}
\mathcal{M}_1 &= |\mathcal{E}|^{-1}\sum_{ij\in \mathcal{E}} \Big[ \frac{N^{-1} \sum_{\tau=1}^N [p_{ij}(\tau) -\hat{p}_{ij}(\tau)]^2}{\sigma_{ij}^2}\Big]^{1/2} \, , \\
\mathcal{M}_2 &= \bigg[|\mathcal{E}|^{-1} \sum_{ij\in \mathcal{E}}\frac{N^{-1} \sum_{\tau=1}^N [p_{ij}(\tau) -\hat{p}_{ij}(\tau)]^2}{\sigma_{ij}^2}\bigg]^{1/2} \, . 
\end{align}
For the data shown in Fig.~\ref{fig:Swiss_flows}, we obtain 
$\mathcal{M}_1 = 0.06862$ and  
$\mathcal{M}_2 = 0.01515$.




Despite the rather small number of observations in the training datasets, we see that the method accurately estimate the state of the system for this real-world test case.  We expect that our results would be of even better quality, if we had access to datasets with observation size comparable to the synthetic data treated in Sec.~\ref{sec:synthetic}.

\section{Predicting Generators}

\begin{figure}[ht!]
\centerline{\includegraphics[width=0.9\columnwidth]{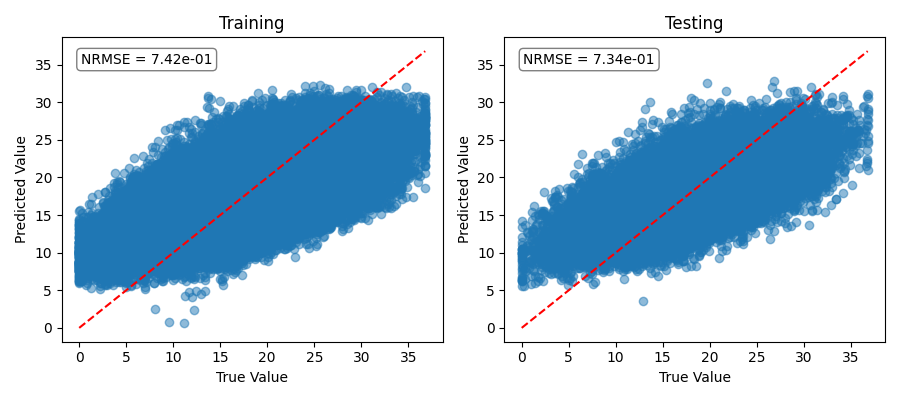}}
\caption{Predicting generators. Model performance for the a) training and b) testing data sets for the German grid, where the single largest generator is left out of the model. }
\label{fig:predict_gen}
\end{figure}

We finally turn to predicting the generators.  Figure~\ref{fig:predict_gen} shows the model performance when only a single generator is left out of the synthetic data set for the German grid.  The model performance is poor, only slightly better than random guessing of the generator value.

We attribute this behavior to the fact that, unlike loads, generators are not necessarily correlated for at least four reasons. First, they come in different types, ranging from very flexible (such as dam hydroelectric plants), to dispatchable with ramp-up and -down times (from less to more constrained: gas, coal and nuclear power plants), to undispatchable (run-of-river hydroelectric plants, solar photovoltaic plants and wind turbines). These varying degrees of flexibility explain why generators sometimes do not produce in situations where they did produce earlier. Second, generation is directly subjected to market rules and generators produce according to a mixture of short-, medium- and long-term contracts. Different operators have different business plans, which directly translate into different production profiles. Third, grid operators have a direct impact on generation to prevent potential congestion. Fourth, many generators are optimized to produce close to their rated power. Therefore, their production profiles are not smooth, continuous curves, but look more like step functions.

These four aspects reduce correlations between the generation profiles of different plants, which explains why our relatively simple linear regression method performs poorly for generators. However, one should keep in mind that almost all generators connected to high- and extra-high-voltage transmission grids have large rated capacities typically exceeding 100 MW. Moreover, there are typically ten times fewer generator than load nodes in a transmission grid. For these reasons, there are often redundant and more secure communication lines between generators and grid operators, reducing the frequency of events with corrupted generation data.

\section{Discussion} \label{sec:discussion}

We construct a novel, data-based method for state-estimation in high voltage transmission grids. The method is based on a simple linear regression algorithm,
which infers control variables instead of state variables.  Essentially all earlier state estimation methods rely on the state variables. Numerical results on realistic synthetic data~\cite{Gil24,Gil25} for three different high voltage transmission grids demonstrated the accuracy of the method. Moreover, for large enough training datasets, we are able to reconstruct close to half the total number of loads with only a small, tolerable decrease in accuracy [see Fig.~\ref{fig:performance_n_loads}] for the larger Spanish and German grids. The ratio of inferrable loads is however significantly smaller for the smaller grid of Switzerland, where only few simultaneously missing loads can be accurately predicted. This is likely so for statistical reasons.  However, specific details of the Swiss power grid, such as larger inhomogeneities in load distributions at different nodes, may also play a role.

The method is further validated on real-life historical datasets for the Swiss power grid. Despite the small volume of data, state estimation works rather well there, with deviations between real and predicted values that are larger by one order of magnitude at most, but still with NRMSE on the order of 10$^{-1}$. It is very likely that a larger training dataset will bring this value down to the level of those obtained with the synthetic dataset.

All other quantities of interest for system operators, such as voltage amplitudes and phases, or current and power flows on lines can be numerically inferred from the inferred power, \textit{i.e.}, control variables, using standard Newton-Raphson solvers. We show that the method reconstructs power flows quite accurately for the real Swiss power grid dataset.

There are two aspects that deserve further investigation. First, we numerically determined which loads are more important in predicting the left-out loads. This information is directly encoded in the elements of the weight matrix $\mathbf{W}^\ell$ defined in Eq.~\eqref{eq:full_model}. We find very different weight distributions [see Fig.~\ref{fig:combined_weights_histogram}] for the different grids we considered. In particular, the distribution is Lorentzian for the Swiss and Spanish grids, while it is Gaussian for the German grid, with additionally very different distribution widths. We determined that this discrepancy is real, \textit{i.e.}, not due to the fitting procedure described in Section~\ref{section:model}, nor to the precise choice of the ridge regression parameter $\alpha$. Beyond that, we cannot explain this observation. Second, it is unclear to us why the method performs so well for loads, but fails so dramatically for generators. 
We should however keep in mind that all state estimation approaches suffer from some shortcomings. We think that, despite these two points, our state estimation method seems promising for deployment working in parallel with other power grid prediction and modeling approaches. Our method adds one new complementary method to the suite of simulation tools. 

We finally mention how our model can be used in practice. In one scenario, a grid operator loses communication with one or more node measurement system(s). Using the database of historical data, a model is trained to infer the missing data. Training the model takes a few seconds of computer time as discussed in Sec.~\ref{sec:synthetic}.
For security applications, one model for each load can be trained and used to predict its behavior.  Deviation from the expected behavior can indicate a node failure, or a hacking or spoofing attack on the grid.

\bibliography{State_Eval_PGrid_ML}
\bibliographystyle{IEEEtran}

\begin{IEEEbiography}[{\includegraphics[width=1in,height=1.25in,clip,keepaspectratio]{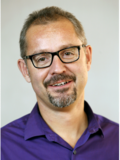}}]{Philippe Jacquod} (Member, IEEE) received the Diplom degree in theoretical physics from the ETHZ, Z\"{u}rich, Switzerland, in 1992, and the PhD degree in
    natural sciences from the University of Neuch\^{a}tel,
    Switzerland, in 1997. He is a professor with the Department of Quantum Matter Physics, University of Geneva, Switzerland, and with the engineering department,
    University of Applied Sciences of Western Switzerland. From 2003 to 2005, he was an assistant professor with the
    theoretical physics department, University of Geneva,
    and from 2005 to 2013, he was a professor
    with the physics department, University of Arizona, Tucson, USA. His main current research interests are in dynamical aspects of power systems and in population dynamics in theoretical ecology. He has published more than 100
    papers in international journals, books and conference proceedings.
\end{IEEEbiography}

\begin{IEEEbiography}[{\includegraphics[width=1in,height=1.25in,clip,keepaspectratio]{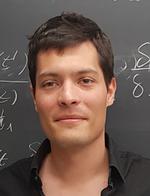}}]
    {Laurent Pagnier} (Member, IEEE) received the M.S. and Ph.D. degrees in theoretical physics from EPFL, Lausanne, Switzerland, in 2014 and 2019, respectively. He is currently a Research Assistant Professor at the University of Arizona. His research interests include the development of novel modeling and monitoring methods for power systems. He is particularly interested in applying machine learning techniques to power systems, with an emphasis on improving the interpretability and trustworthiness of data-driven methods to facilitate their adoption within the power systems community.
\end{IEEEbiography}

\begin{IEEEbiography}[{\includegraphics[width=1in,height=1.25in,clip,keepaspectratio]{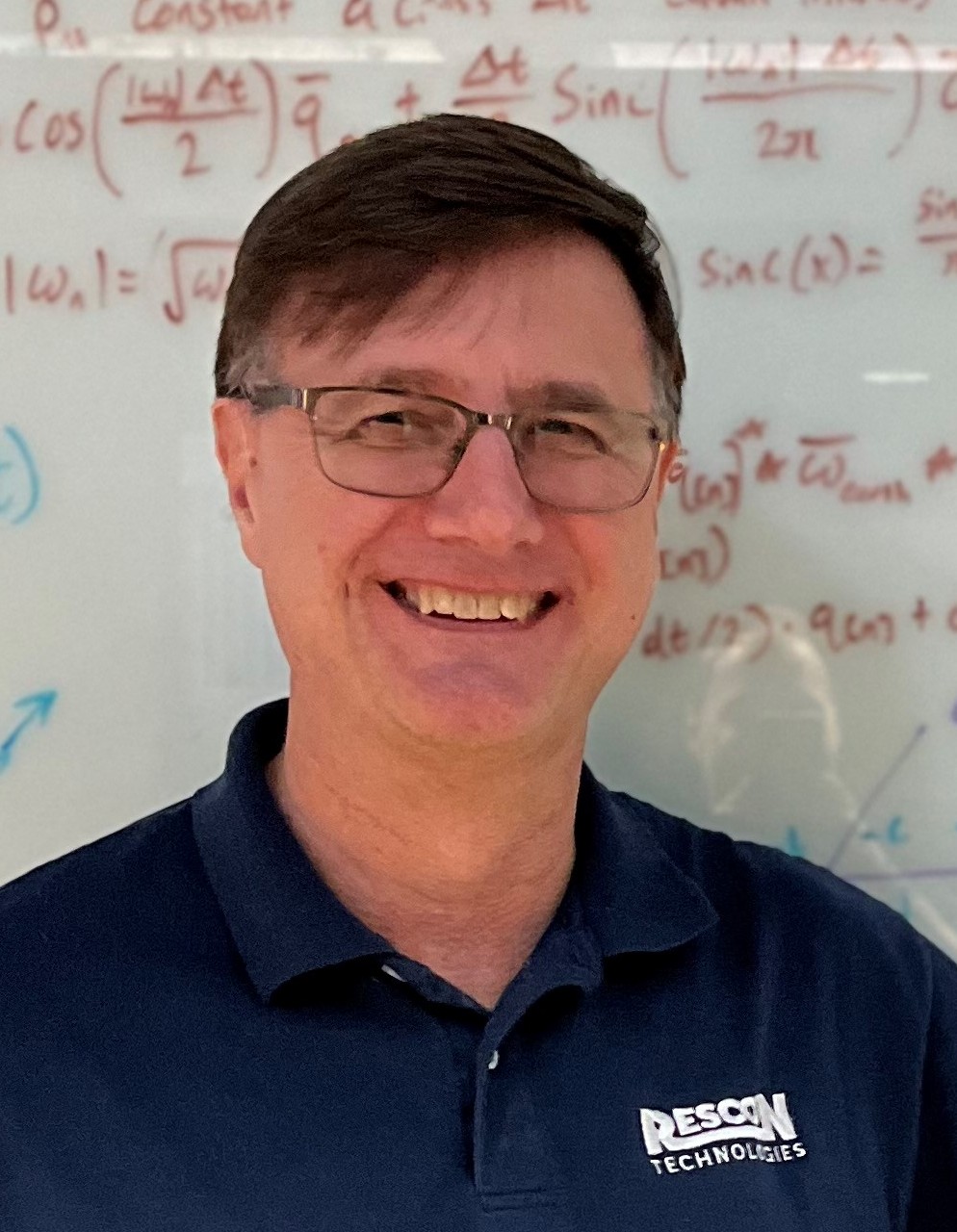}}]
    {Daniel J. Gauthier} is co-founder and chief technology officer of ResCon Technologies.  He received the B.S., M.S., and Ph.D. degrees in optics from the University of Rochester, Rochester, NY, USA in 1982, 1983, and 1989, respectively.  He was a post-doctoral researcher at the University of Oregon, OR, USA, from 1989-1991, a professor of physics at Duke University, Durham, NC, USA, from 1992-2015, and a professor of physics at The Ohio State University, Columbus, OH, USA from 2016-2024. He is commercializing machine learning algorithms to solve industry-relevant problems. He is a Fellow of Optica and the American Physical Society.
\end{IEEEbiography}

\EOD

\end{document}